\documentclass[trackchanges]{aastex701}

\usepackage[export]{adjustbox}

\begin{document}

\title{Observations of a Faint Nonthermal Onset before a GOES C-class Flare}

\author[0000-0003-1652-6835]{Natália Bajnoková}
\affiliation{School of Physics \& Astronomy, University of Glasgow, University Avenue, Glasgow G12 8QQ, UK}
\email{n.bajnokova.1@research.gla.ac.uk}

\author{Iain G. Hannah}
\affiliation{School of Physics \& Astronomy, University of Glasgow, University Avenue, Glasgow G12 8QQ, UK}
\email{Iain.Hannah@glasgow.ac.uk}

\author{Hannah Collier}
\affiliation{University of Applied Sciences and Arts Northwestern Switzerland (FHNW), Bahnhofstrasse 6, 5210 Windisch, Switzerland}
\affiliation{ETH Zürich, Rämistrasse 101, 8092 Zürich Switzerland}
\email{hannah.collier@fhnw.ch}

\author{Stephen M. White}
\affiliation{Air Force Research Laboratory, Albuquerque, NM, USA}
\email{stephen.white.24@us.af.mil}

\author{Lindsay Glesener}
\affiliation{School of Physics \& Astronomy, University of Minnesota Twin Cities, Minneapolis, MN 55455, USA}
\email{glesener@umn.edu}

\author{Reed B. Masek}
\affiliation{School of Physics \& Astronomy, University of Minnesota Twin Cities, Minneapolis, MN 55455, USA}
\email{masek014@umn.edu}

\author{Marianne S. Peterson}
\affiliation{School of Physics \& Astronomy, University of Minnesota Twin Cities, Minneapolis, MN 55455, USA}
\email{pet00184@umn.edu}

\author{Säm Krucker}
\affiliation{University of Applied Sciences and Arts Northwestern Switzerland (FHNW), Bahnhofstrasse 6, 5210 Windisch, Switzerland}
\affiliation{Space Sciences Laboratory, University of California, Berkeley, CA 94720, USA}
\email{krucker@berkeley.edu}

\author{Hugh S. Hudson}
\affiliation{School of Physics \& Astronomy, University of Glasgow, University Avenue, Glasgow G12 8QQ, UK}
\affiliation{Space Sciences Laboratory, University of California, Berkeley, CA 94720, USA}
\email{Hugh.Hudson@glasgow.ac.uk}

\begin{abstract}

We present analysis of a GOES C1-class flare from 2022 September 6, which was jointly observed as occulted by Nuclear Spectroscopic Telescope ARray (NuSTAR) and on-disk by Spectrometer/Telescope for Imaging X-rays (STIX). NuSTAR observed faint coronal nonthermal emission as well as plasma heating $>$ 10 MK, starting 7 minutes prior to the flare. This onset emission implies that during this time, there is a continuous electron acceleration in the corona which could also be responsible for the observed heating. The nonthermal model parameters remained consistent throughout the entire onset, indicating that the electron acceleration process persisted during this time. Furthermore, the onset coincided with a series of type III radio bursts observed by Long Wavelength Array-1, further supporting the presence of electron acceleration before the flare began. We also performed spectral analysis of the impulsive flare emission with STIX (thermal and footpoint emission). STIX footpoints and the onset coronal source were found to have similar electron distribution power-law indices, but with increased low-energy cut-off during the flare time. This could suggest that the nonthermal onset is an early signature of the acceleration mechanism that occurs during the main phase of the flare.

\end{abstract}

\keywords{\uat{Active sun}{18} --- \uat{Solar atmosphere}{1477} --- \uat{Solar corona}{1483}  --- \uat{Solar flares}{1496} --- \uat{Solar X-ray emission}{1536}}

\section{Introduction}\label{sec:intro}

Preflare (or onset) emission \citep[e.g.][]{Benz_radio_onset, Battaglia_2009} and flare precursors \citep[e.g.][]{Farnik_2003, Wang_2017} have been extensively studied in the hope of better describing the physical processes that lead up to initial flare energy release. In particular, soft (SXR) and hard (HXR) X-ray observations are crucial for studying the preflare emission, as they allow us to directly probe plasma heating and electron acceleration in the solar corona \citep{Krucker_coronal_review}. 

The characteristic increase in SXR emission prior to the impulsive
phase has now been identified as a ``hot onset precursor event" (HOPE), with properties distinctly different from the common description of ``preheating" for the precursor phase \citep{Hugh_onset}. During this phase, the soft X-ray spectrum immediately appears at high temperature ($\sim$10 MK) which does not vary appreciably as the emission measure steadily increases. A statistical study of GOES flares in 2010-2011 showed that 75\% had hot-onset precursors \citep{Silve_onset}, and \citet{2025SoPh..300....2H} suggests that virtually all flares have this property. Studies of the hot-onsets with the Solar Orbiter - Spectrometer/Telescope for Imaging X-rays \citep[STIX;][]{Krucker_2020} show a similar trend, often exhibiting elevated plasma temperatures even earlier than GOES \citep{Andrea_onset}. Sustaining such elevated plasma temperatures, prior to the main energy release, requires continuous energy input \citep{Battaglia_2009}. Furthermore, observations from \citet{Hugh_onset} and \citet{Andrea_onset} suggest that the onset SXR emission probably originates from compact hot plasma regions near the chromosphere around the footpoints and low-lying loops; however, no concrete evidence of the heating mechanism or nonthermal emission has been identified yet.

In this Letter, we analyze the onset HXR emission of a C-class flare observed by the Nuclear Spectroscopic Telescope ARray \citep[NuSTAR;][]{Harrison_2013} and its link to the HXR emission during the flare observed by STIX. Section \ref{sec:overview} provides an overview of the observed flare. In Section \ref{sec:onset} we present the spectral fitting and imaging results from the preflare and onset time ranges observed by NuSTAR and discuss them in the context of EUV emission observed by the Solar Dynamics Observatory’s Atmospheric Imaging Assembly \citep[AIA;][]{AIA_paper}. Finally, in Section \ref{sec:flare} we discuss the imaging and spectral fitting results of the flare from STIX.

\section{Observation Overview} \label{sec:overview}

\begin{figure}[ht!]
    \begin{minipage}{0.68\textwidth}
        \includegraphics[width=1\textwidth]{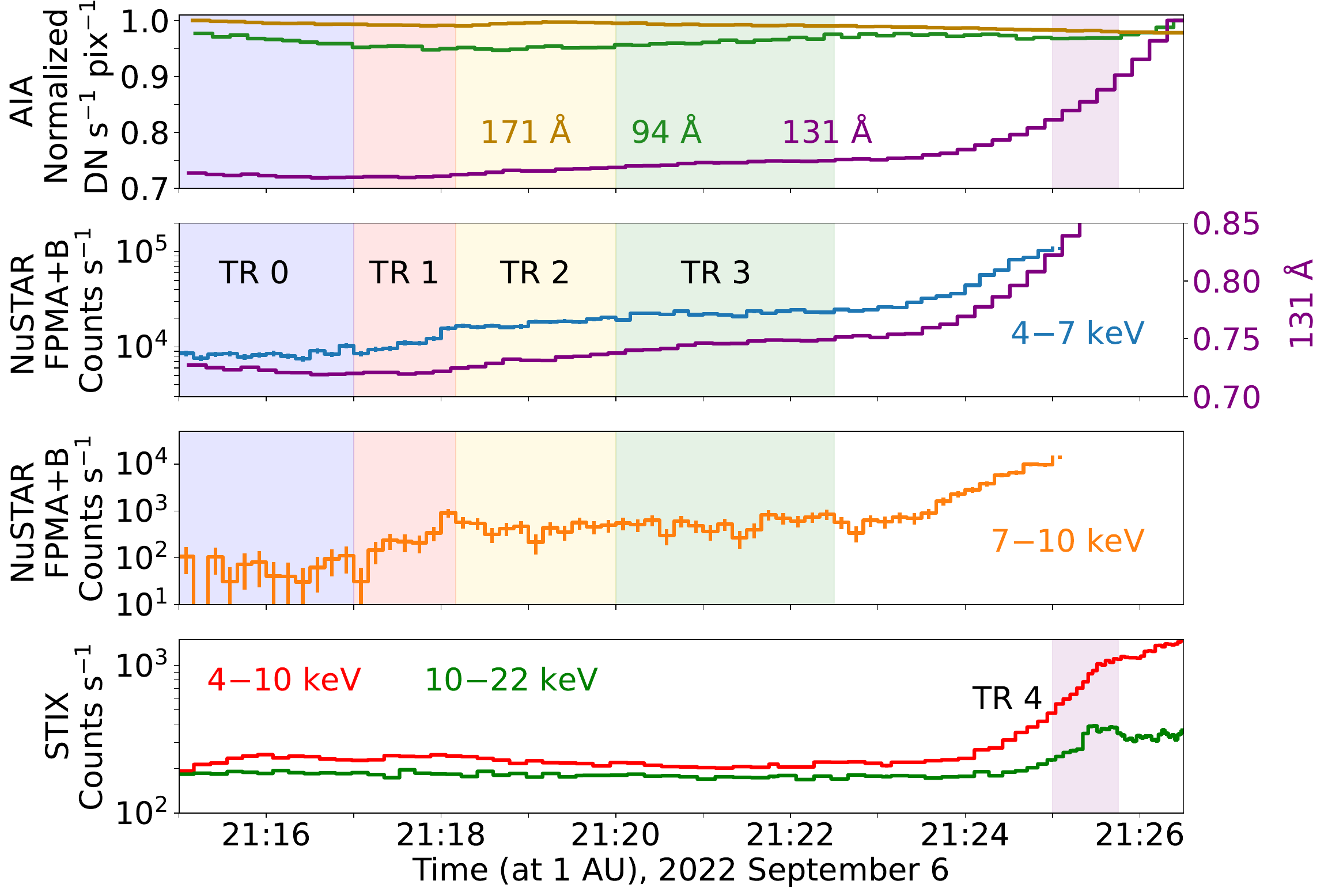}
    \end{minipage}
    \begin{minipage}{0.31\textwidth}
        \includegraphics[width=0.95\textwidth]{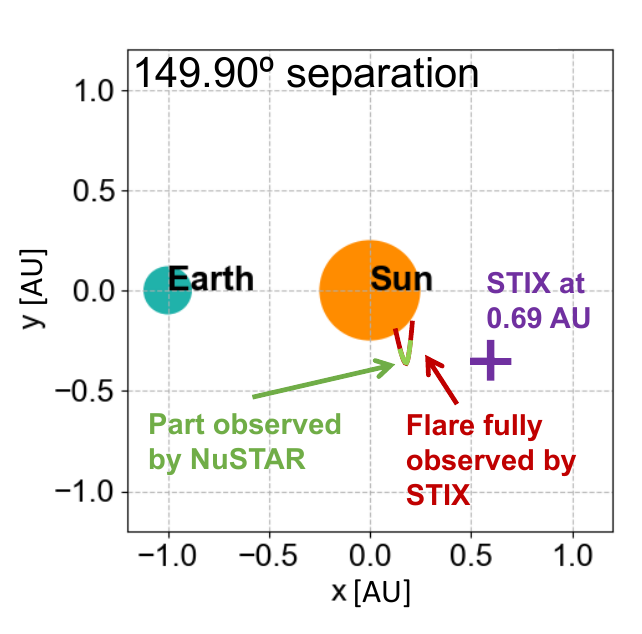}

        \includegraphics[width=1\textwidth]{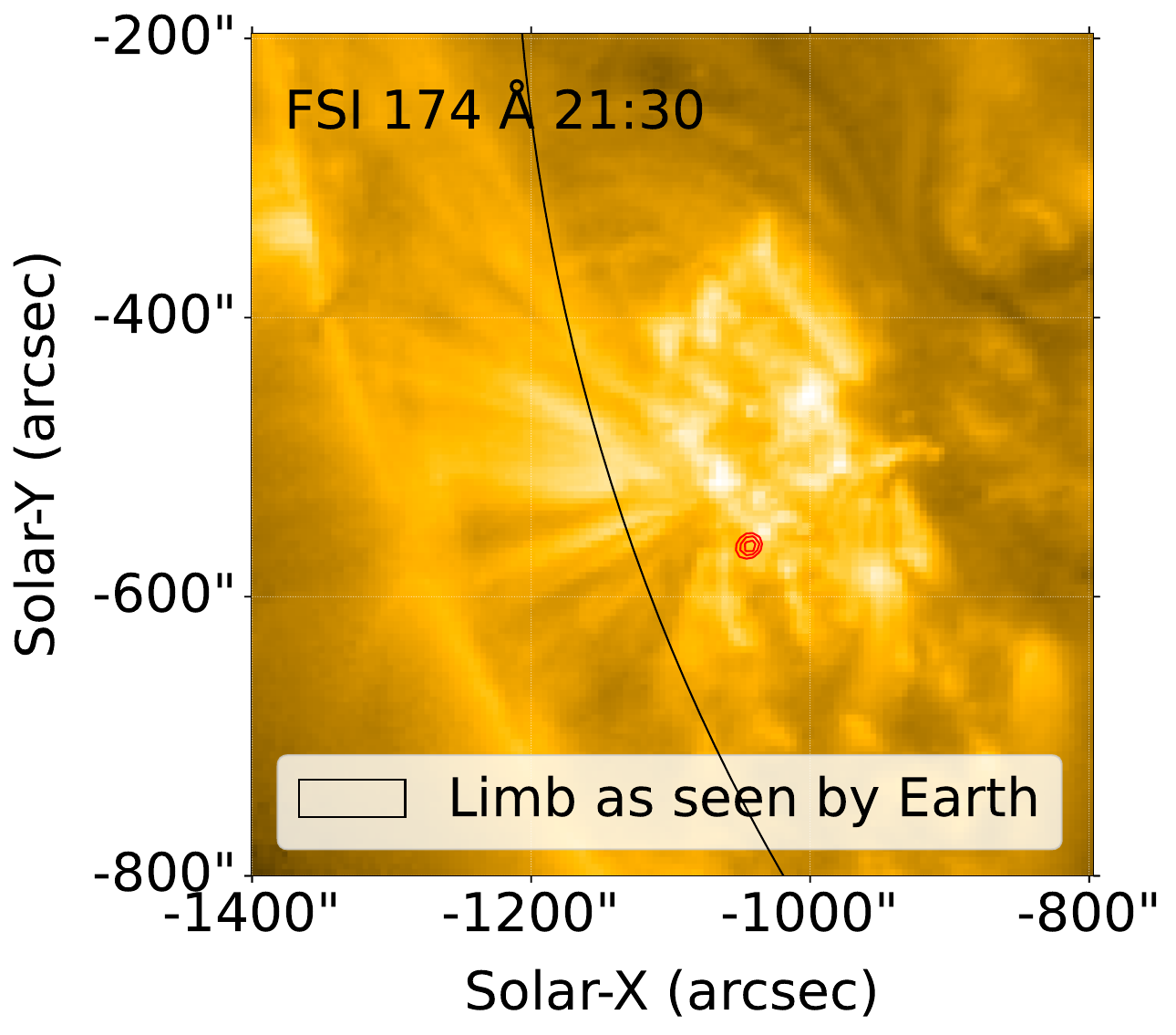}
    \end{minipage}
        
    \caption{(Left) AIA, NuSTAR 4-7 keV (including 131 \AA{} AIA lightcurve from the above panel but with y-aixs limited to 0.70 - 0.85 range), NuSTAR 7-10 keV and STIX (time corrected to 1 AU) lightcurves from the 2022 September 6 observation campaign. The time ranges used for integrating the spectra are highlighted in colour shaded regions. (Top right) The NuSTAR (Earth)--STIX--flare positioning for the observed events. The separation angle defines the approximate Earth--Sun--STIX separation. (Bottom right) EUI/FSI 174 \AA{} image from the flare decay time. The red contours indicates the approximate flare location as seen by STIX and the black line indicates the limb seen from Earth. 
    \label{fig:lightcurves}}
\end{figure}

On 6 September 2022 at 21:31 UTC, NuSTAR and STIX jointly detected a GOES C1-class flare (as seen from Earth) that was observed to be occulted by the west solar limb from Earth and on disk for Solar Orbiter, as shown in the spacecraft configuration in the top right panel in Figure \ref{fig:lightcurves}. The STIX flare location overlain on the Full Sun Imager (FSI) on board Solar Orbiter’s Extreme Ultraviolet Imager (EUI) \citep[][]{EUI_paper} image, shown in the bottom right panel of Figure \ref{fig:lightcurves}, suggests that the flare was $\sim$ 7.4\textdegree{} behind the limb for NuSTAR. This is an angle estimate for the center of emission, and the whole STIX source covers 6.6\textdegree{} to 8.2\textdegree{} from the limb. The AIA, NuSTAR, and STIX time profiles from the observations are shown in Figure \ref{fig:lightcurves}.

NuSTAR lightcurves, summed over its two quasi-identical telescopes, FPMA and FPMB, show an increase in HXR emission starting $\sim 7$ minutes before the flare. For analysis, we divide these into three time ranges (TRs). NuSTAR first observed a steep increase in emission (especially prominent in the 7-10 keV energy range) from 21:17-21:18 (TR 1). This is followed by a slow increase in the emission from 21:18 - 21:22, which is mainly visible in the 4-7 keV energy range. For spectral analysis, we split this part into two TRs:
21:18 - 21:20 (TR 2) and 21:20 - 21:22 (TR 3). Due to the low livetime ($<$ 0.3\%) during these observations, the NuSTAR spectra had to be corrected for gain and pile-up, detection of multiple photons as a single higher-energy photon \citep{grefenstette2016}. The approach used is described in \citet{NuSTAR_STIX_fitting}, where leveraging NuSTAR’s pixel detection grading system, we account for pile-up by simultaneously fitting a pile-up model to the Grade 21–24 spectra (a direct measure of double pixel detections) whilst also applying it and fitting the physical model to the Grade 0–4 spectra (single and double pixel detections). Unfortunately, because of the extremely high flux from the flaring time, NuSTAR spectral data are only available until $\sim$ 21:22. After that time, it is not possible to obtain reliable spectral fits due to extreme pile-up and eventually on-board event rejection. STIX did not observe the onset activity, as it is likely to be too faint, and the emission was dominated by the bright, unocculted active region. Furthermore, the STIX X-ray background is elevated for this observation due to a solar energetic particle (SEP) event that occurred several hours before the flare \citep[for more details on SEP contamination of STIX detectors, see][]{STIX_SEP}{}. Therefore, the faint onset emission could be further hidden by the elevated background.

The flare was fully observed by STIX; however, we only analyse the impulsive, nonthermally dominated part of the flare from 21:25:16 to 21:25:45 (TR 4) to investigate the initial electron acceleration. Due to the elevated background from the SEP contamination, the STIX data required a careful background interpolation prior to the spectral fitting.

\section{Analysis of the preflare and onset emission} \label{sec:onset}

In this section, we present spectral fitting and imaging results from the integrated NuSTAR emission from TRs 0-3 highlighted in the NuSTAR lightcurves in Figure \ref{fig:lightcurves}. For this work, we use the spectral fitting routine provided in the SunPy solar X-ray package, sunkit-spex\footnote{\href{https://github.com/sunpy/sunkit-spex}{\nolinkurl{https://github.com/sunpy/sunkit-spex}}}.

\subsection{Spectral fitting}

\begin{figure*}
\centering
\begin{minipage}{0.24\textwidth}
    \includegraphics[width=1\textwidth]{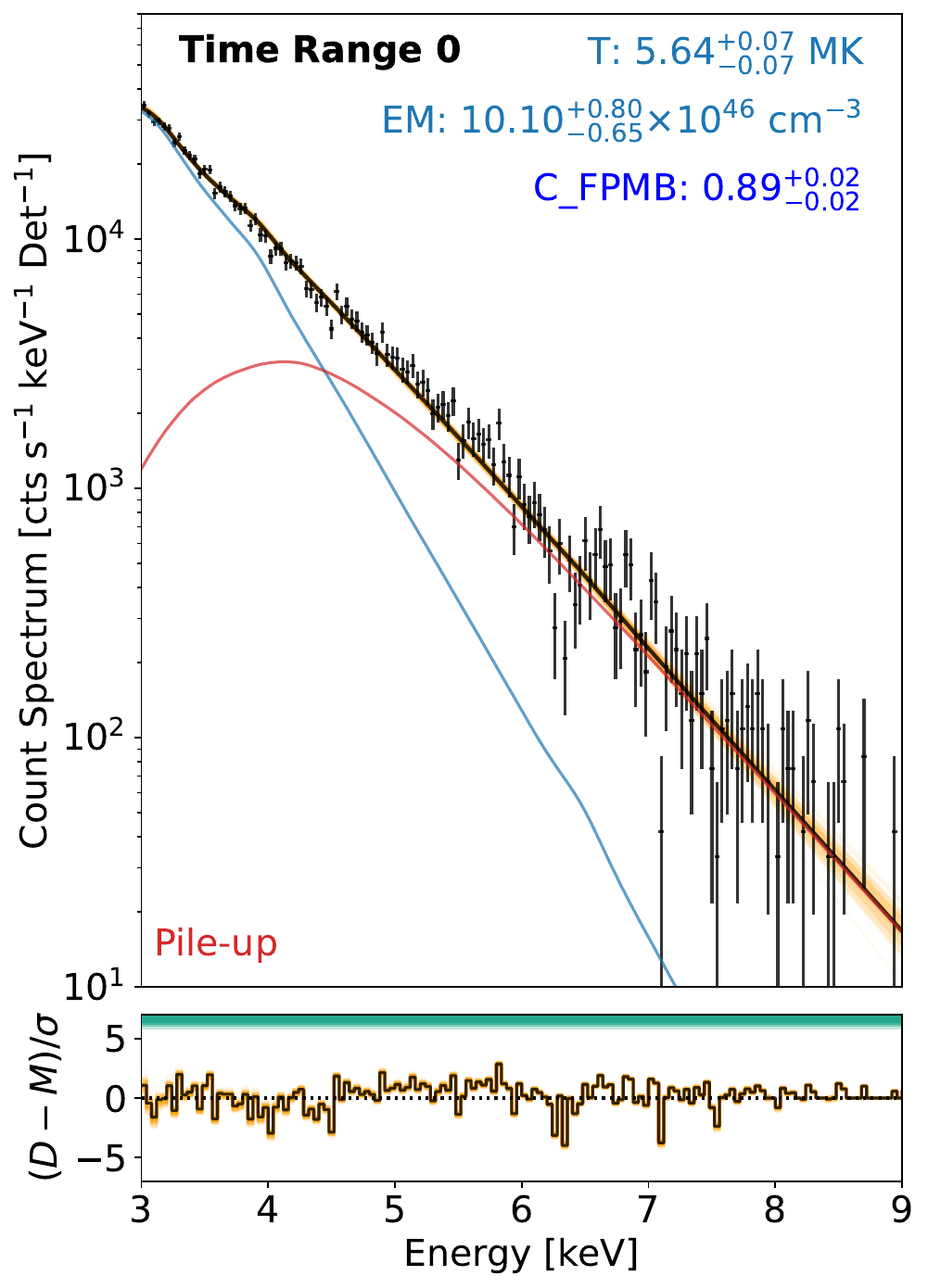}
\end{minipage}
\begin{minipage}{0.24\textwidth}
    \includegraphics[width=1\textwidth]{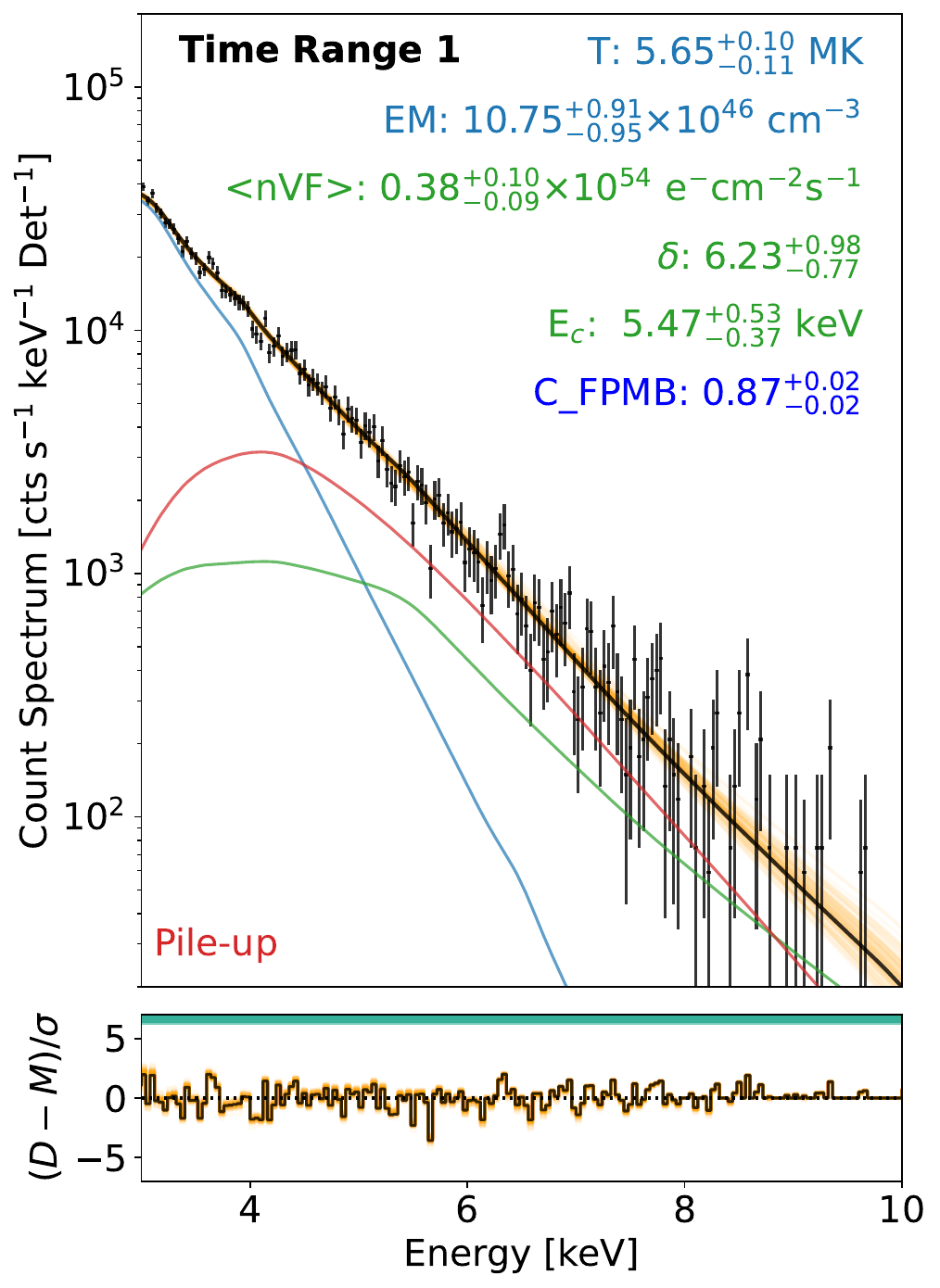}
\end{minipage}
\begin{minipage}{0.24\textwidth}
    \includegraphics[width=1\textwidth]{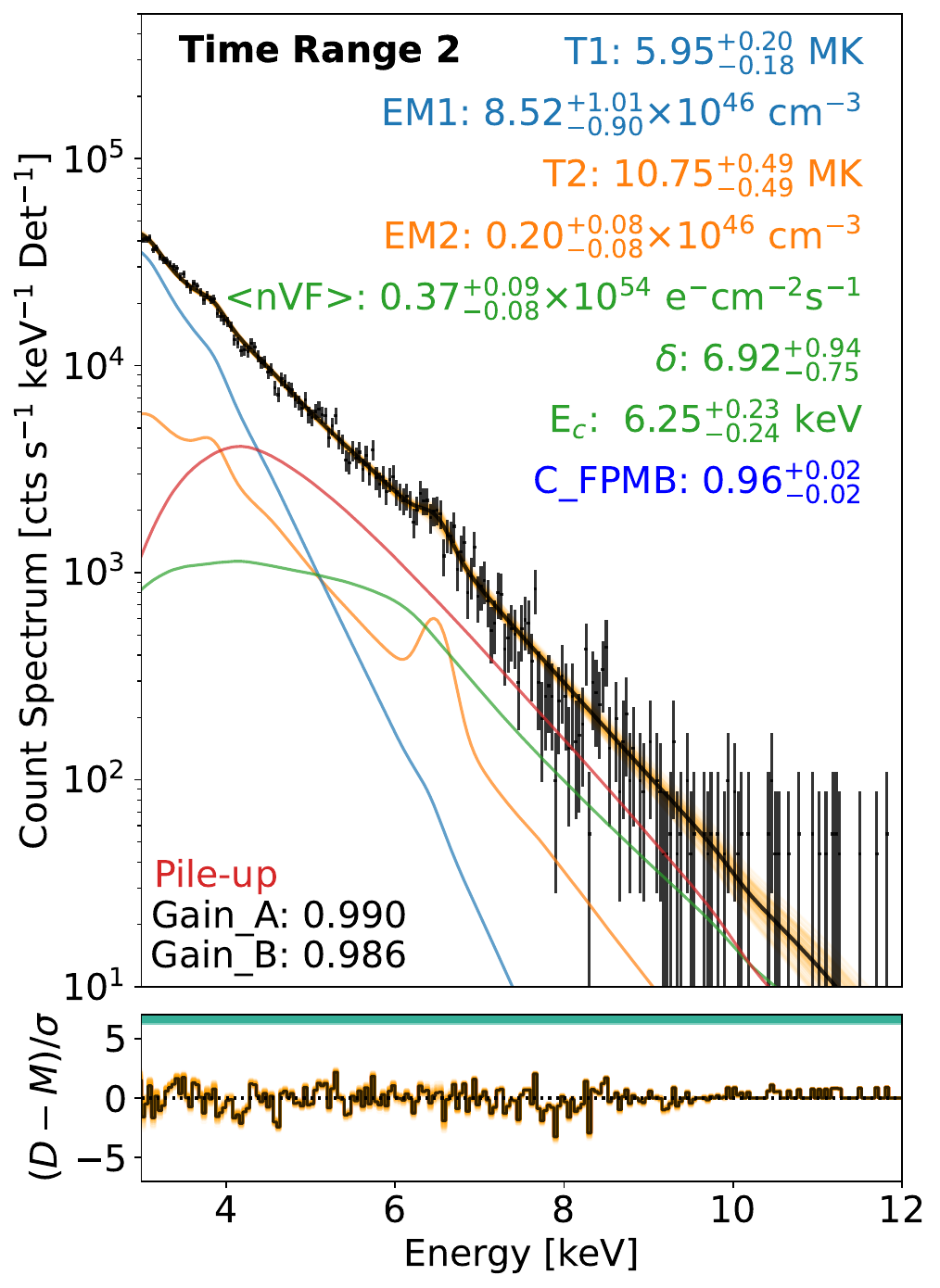}
\end{minipage}
\begin{minipage}{0.24\textwidth}
    \includegraphics[width=1\textwidth]{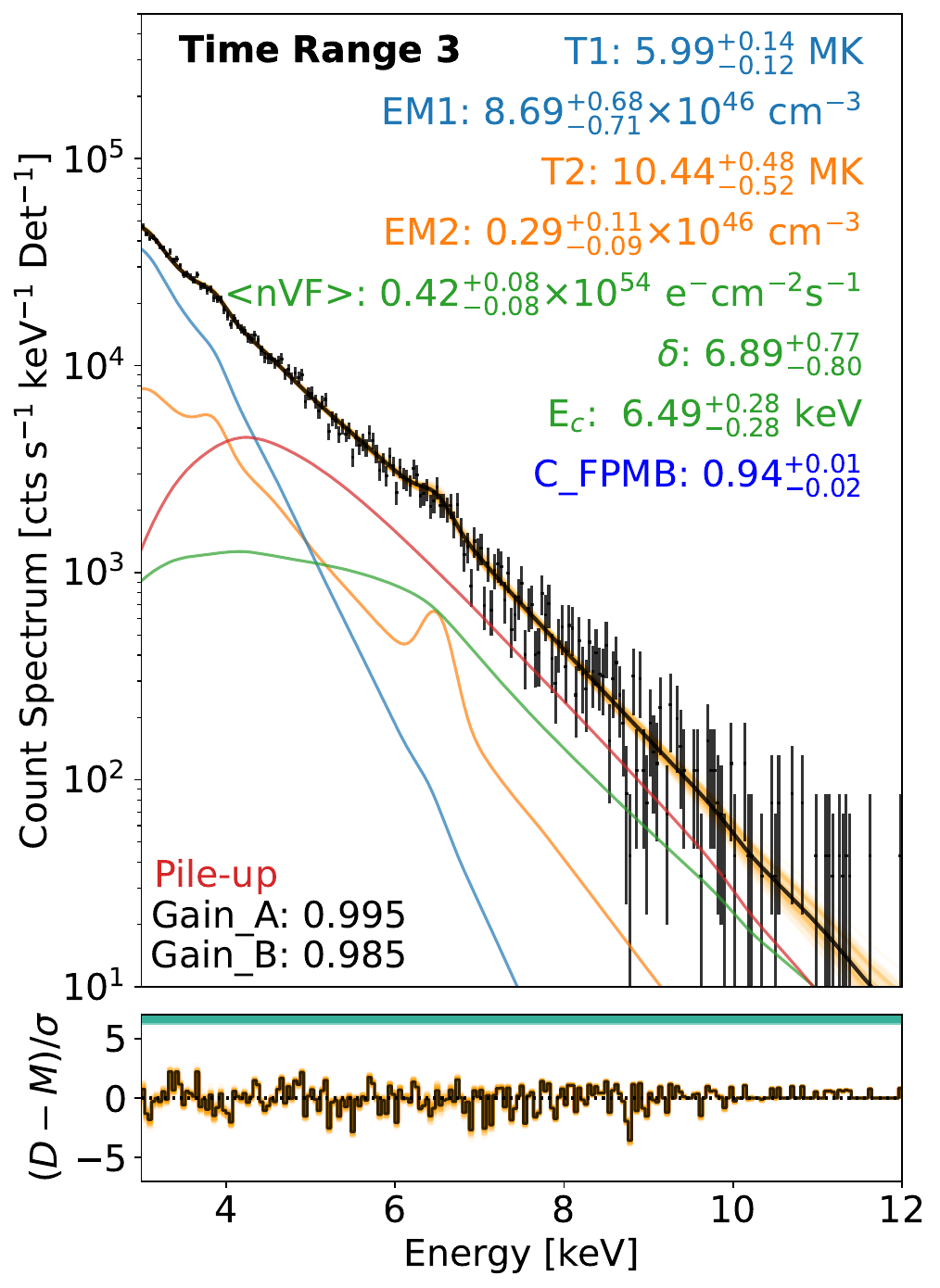}
\end{minipage}

\caption{NuSTAR spectral fits and residuals from time ranges 0-3. The spectral fits show averaged FPMA and B spectra (with the mean pile-up model shown in red), however the fitting was done simultaneously to the separate FPM spectra and with the pile-up spectra as described in \citet{NuSTAR_STIX_fitting}. The models fitted to FPMB spectra were scaled by a factor C$_{\text{FPMB}}$ to account for systematic differences between the two FPMs. The fitted model includes thermal (blue and orange lines) and thin-target nonthermal components (green). The black line indicates the mean MCMC-fitted model with several sample fits shown in yellow. The fitted energy range is shown in the solid green horizontal line.}
\label{fig:nustar_spectra}
\end{figure*}

The NuSTAR spectral fits are shown in Figure \ref{fig:nustar_spectra} and the best-fit parameters are summarised in Table \ref{tab:spec_params}. The preflare spectral fit from TR 0 (first panel in Figure \ref{fig:nustar_spectra}) shows a preflare temperature around 5.64 MK. The beginning of the onset, TR 1 (second panel), was best fitted with a thermal model with a temperature similar to the preflare time, as well as a nonthermal thin-target component, which dominates the spectrum $>6$ keV. We chose to fit the NuSTAR spectra with a thin-target model \citep[][]{Holman_emission}, which assumes negligible electron energy loss, because it is expected that the emission originates in the corona \citep{Krucker_coronal_review}. In such a low density environment, the traversing electrons lose little energy as they interact with ambient particles. The best-fit thin-target model parameters for TR1 were found to be: electron spectral index $\delta=6.23^{+0.98}_{-0.77}$, the low-energy cut-off E$_\textsc{c}=5.47^{+0.53}_{-0.37}$ keV, and mean source electron spectrum $<$nVF$>=$$0.38^{+0.10}_{-0.09}\times 10^{54}$ e$^{-}$cm$^{-2}$s$^{-1}$. We also attempted to fit the spectra with single- and double-thermal models, resulting in lnL values of -368 and -364, respectively. However, the thermal+thin-target model combination provides the best fit, with an lnL of -326.

TRs 2 (third panel) and 3 (fourth panel) show an additional presence of a 10-11 MK thermal model component, along with the lower-temperature preflare component. This would appear to be the hot-onset emission previously detected in flares via GOES observations \citep[][]{Hugh_onset, Silve_onset, Andrea_onset}. This result is consistent with the increase in emission observed in the 131 \AA{} AIA channel, as this channel has a peak in its temperature response at about 10 MK \citep{AIA_response}. While the 131 \AA{} temperature response peaks both below 1 MK and around 10 MK, we conclude that the increase in emission is due to 10 MK plasma, as none of the cooler channels—such as the 171 \AA{} lightcurve shown in the top panel of Figure \ref{fig:lightcurves}—show any significant increase over time. Unfortunately, the GOES lightcurves during the onset time are dominated by emission from another location; therefore, we cannot confirm the hot-onset via this data. Furthermore, both TRs were also fitted with a thin-target nonthermal model, yielding best-fit parameters similar to those of TR 1 (apart from a small increase in the low-energy cutoff, E$_\textsc{c}$). The presence of the nonthermal emission before (TR1) and during the hot onset (TR2 and 3) suggests that this heating $>$ 10 MK, could be due to these nonthermal electrons.

\subsection{Imaging} \label{sec:imaging}

\begin{figure*}
\centering
\begin{minipage}{0.25\textwidth}
\includegraphics[trim={0 1cm 0 0},clip, width=1\textwidth]{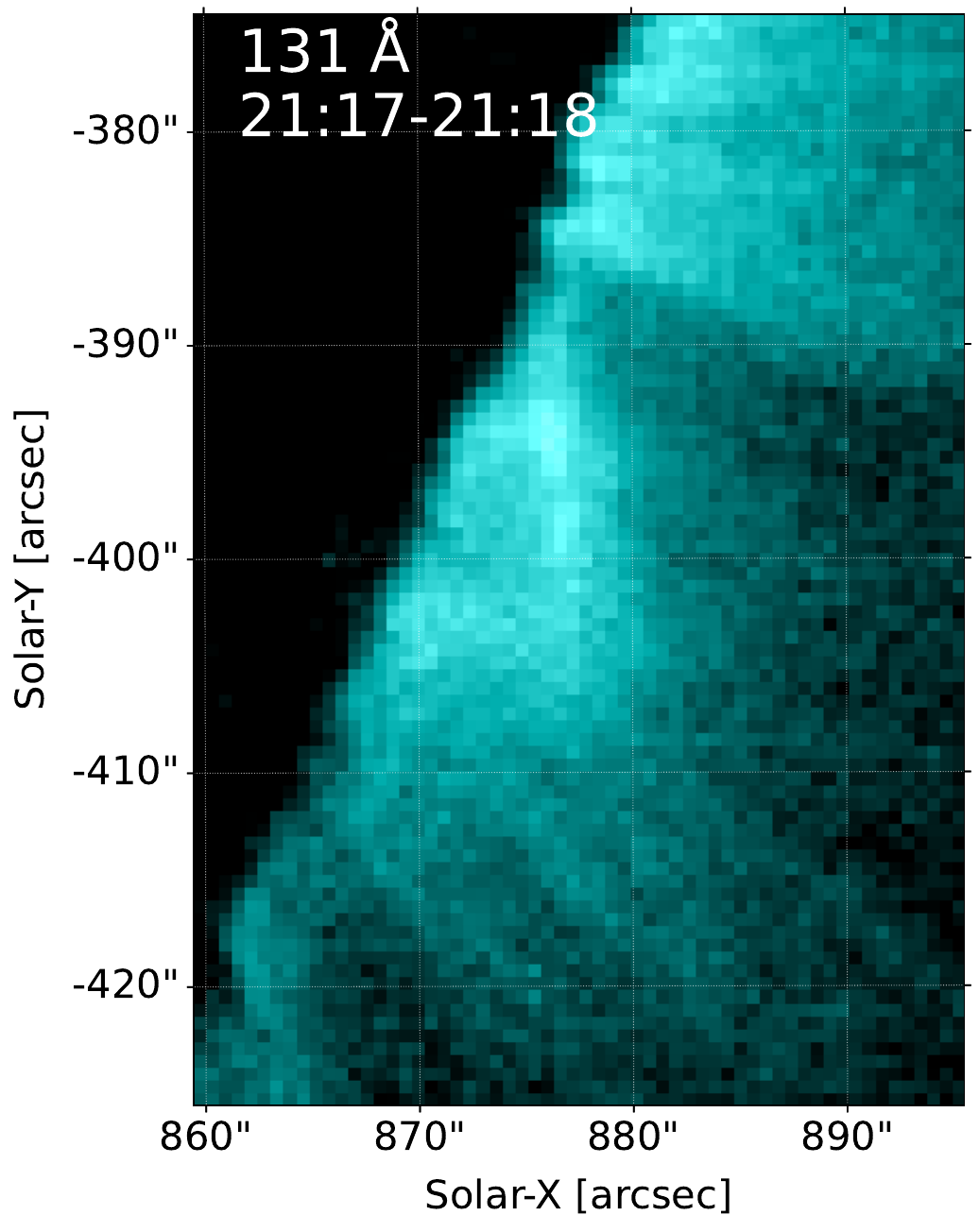}

 \includegraphics[width=1\textwidth]{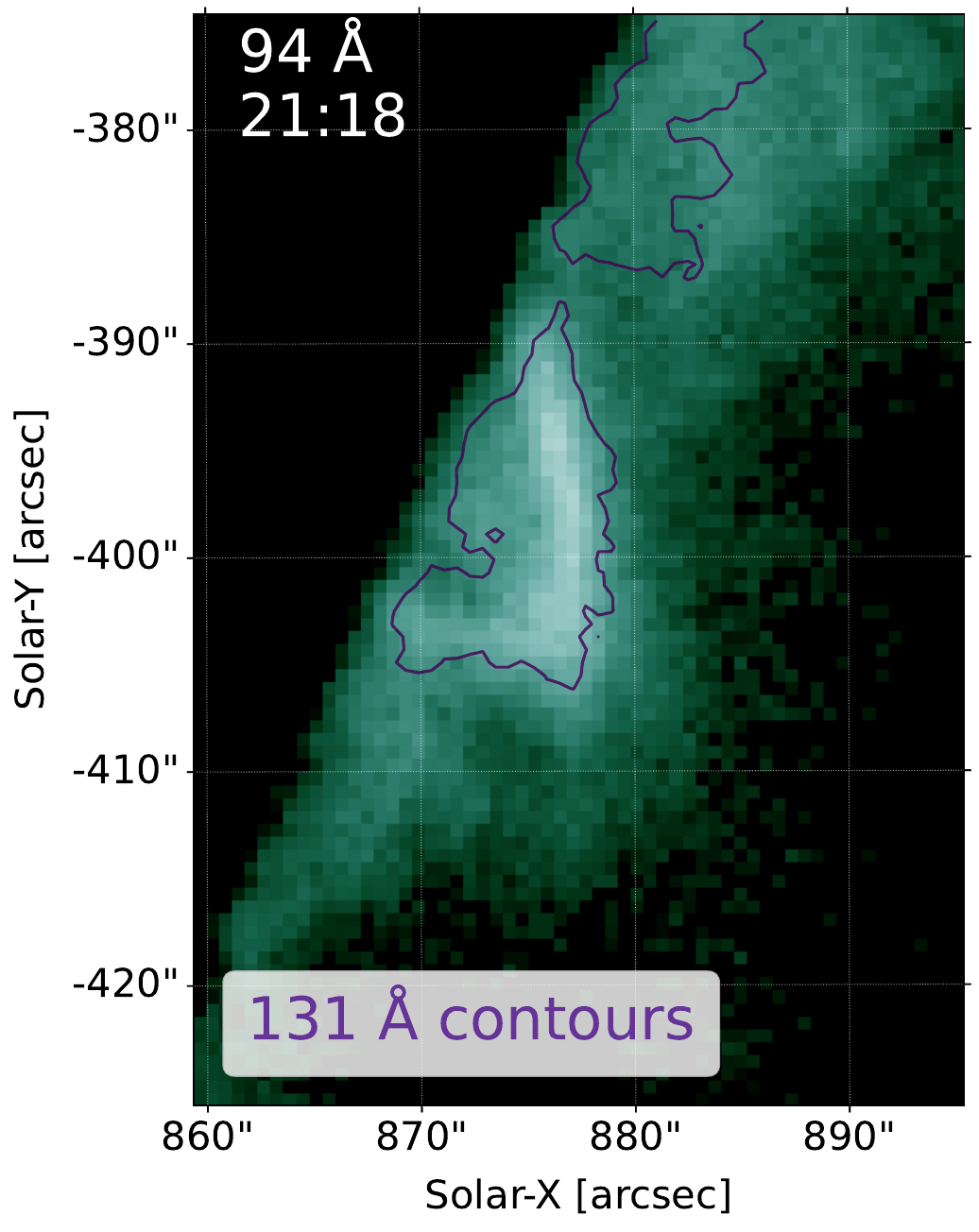}
\end{minipage}
\begin{minipage}{0.25\textwidth}
\centering  
    \includegraphics[trim={1cm 1cm 0 0},clip, width=0.95\textwidth]{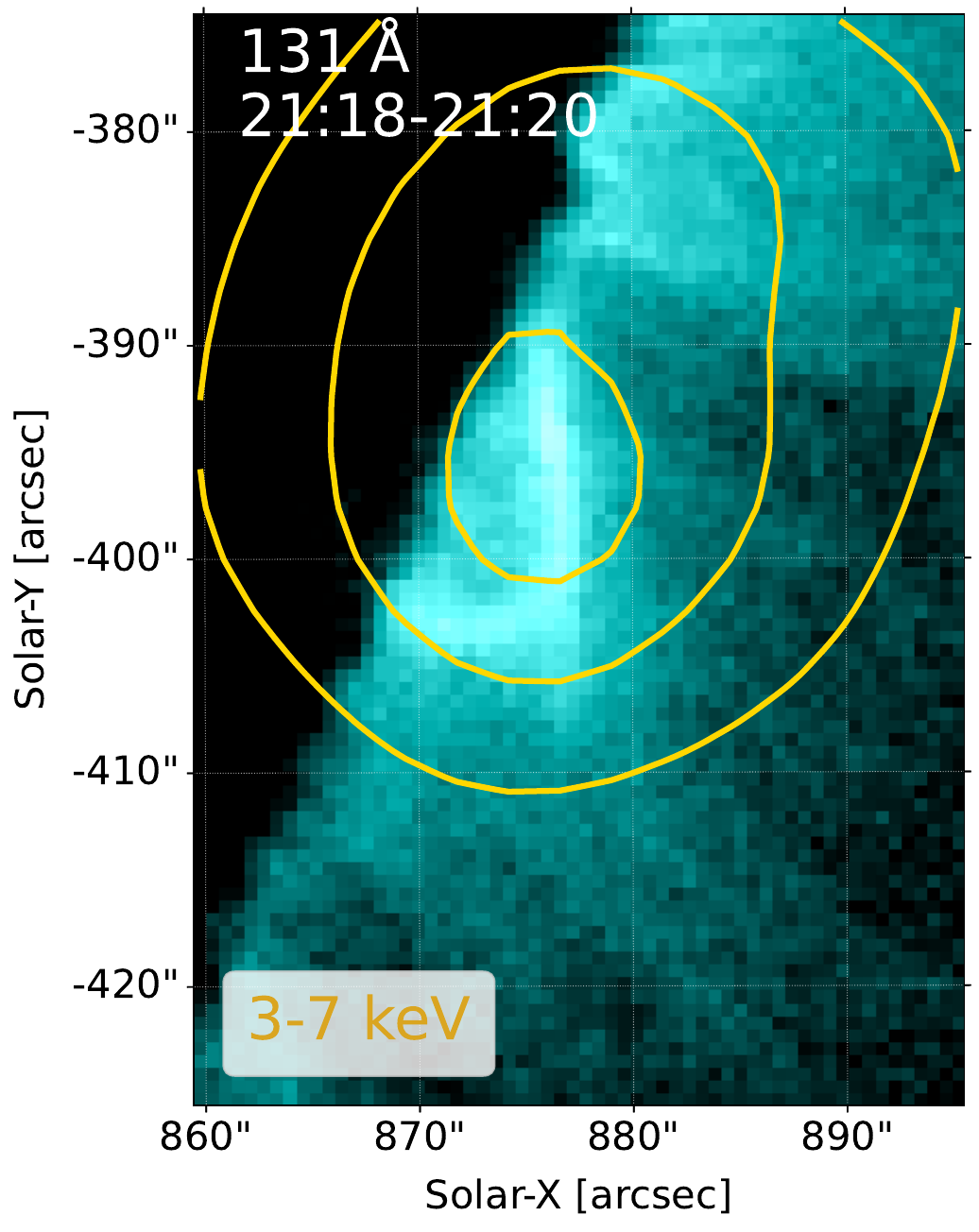}     

    \includegraphics[trim={1cm 0 0 0},clip, width=0.95\textwidth]{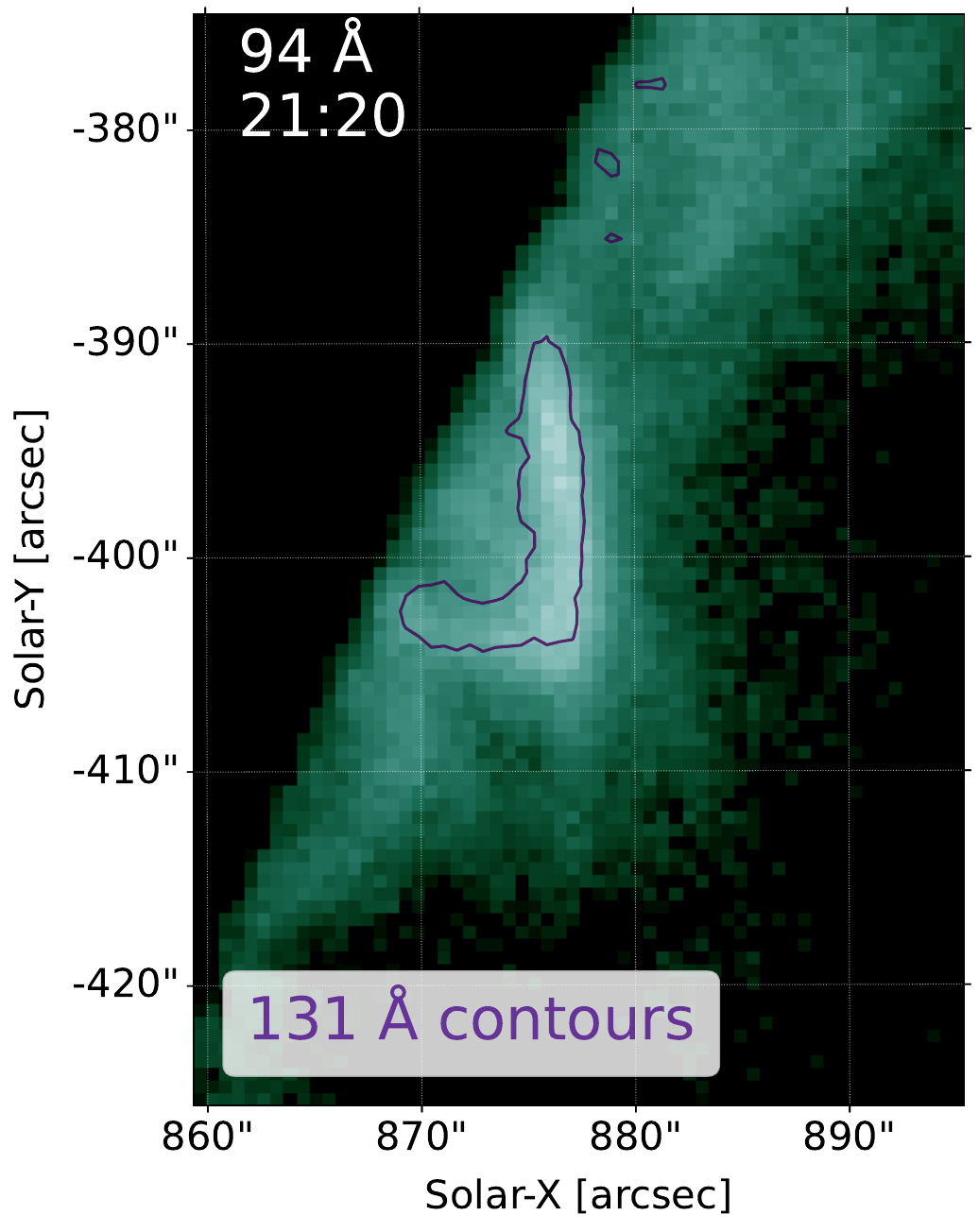}
\end{minipage}
\begin{minipage}{0.25\textwidth}
\centering
    \includegraphics[trim={1cm 1cm 0 0},clip, width=0.95\textwidth]{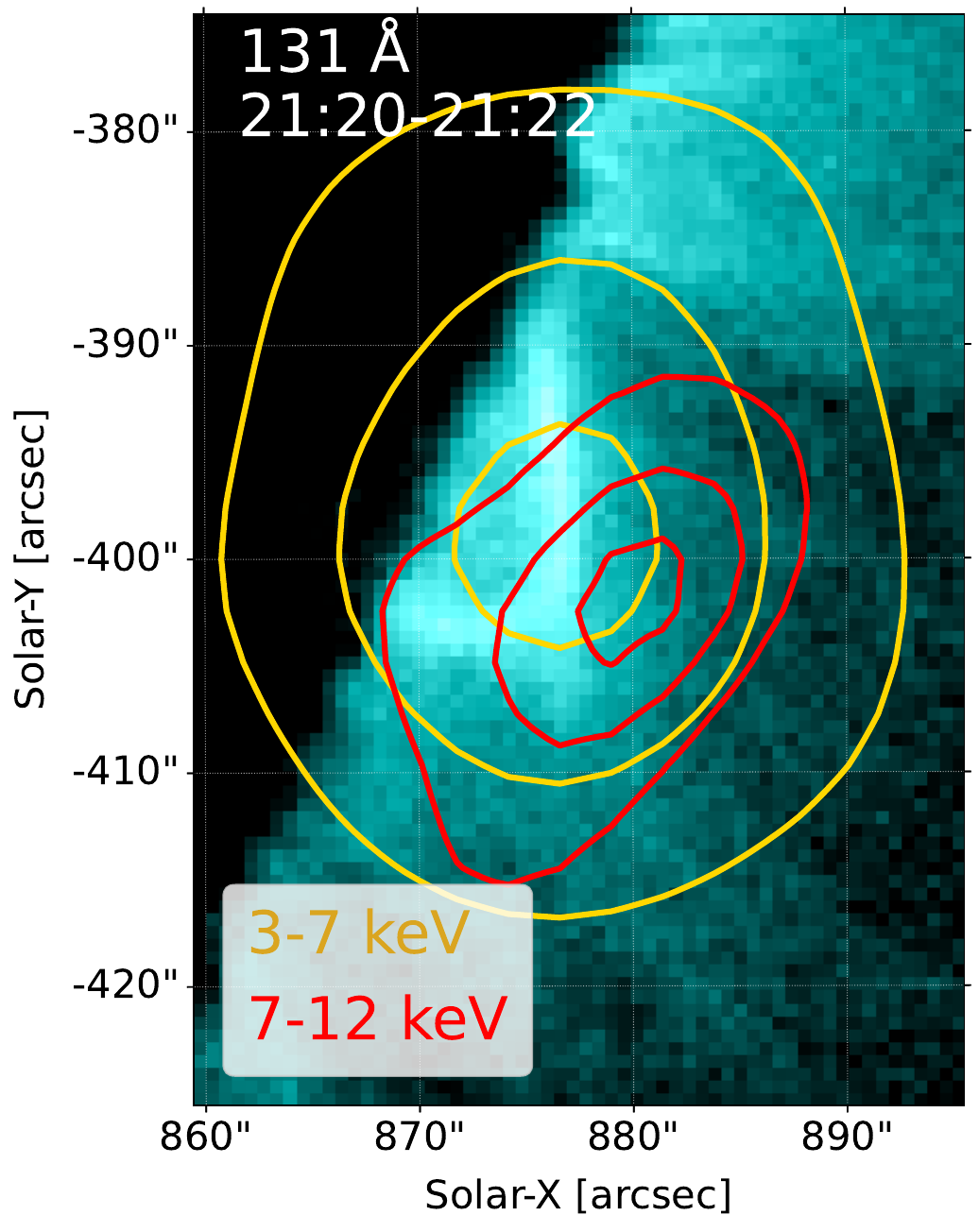}

    \includegraphics[trim={1cm 0 0 0},clip, width=0.95\textwidth]{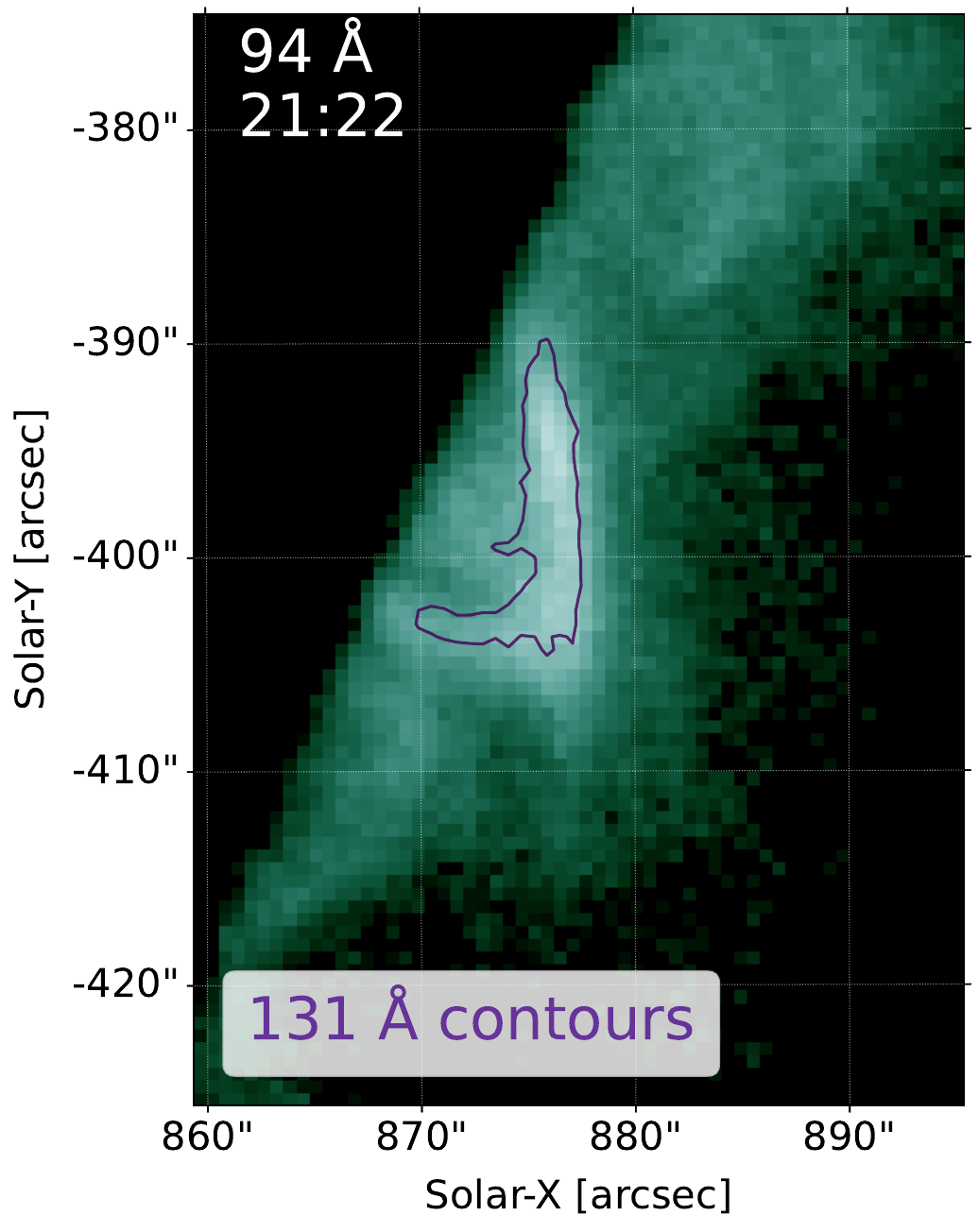}
\end{minipage}
\caption{(Top row) NuSTAR 60, 80, 95 \% contour levels (summed over the two focal plane modules) from TRs 1 (left), 2 (middle) and 3 (right) overlain on 131 \AA{} AIA images. The yellow contours cover thermally dominated energy range and the red contour cover the nonthermally dominated energy range. A Gaussian filter with $\sigma$ of 2 pixels ($\approx 5"$) was used to smooth the NuSTAR images.(Bottom row) 94 \AA{} AIA images (with 60 \% contour levels from the corresponding 131 \AA{} AIA images) from the same time ranges show a loop structure that was present before the flare onset.}
\label{fig:images}
\end{figure*}

Figure \ref{fig:images} shows the NuSTAR contours from the onset time overlayed on 131 \AA{} AIA EUV images. The NuSTAR contours are summed over FPMA and FPMB, with each having an angular resolution of 18 arcsec. We only include single pixel detections (labeled as Grade 0 in the NuSTAR processing software) to eliminate any effects of pile-up. By removing all multi-pixel detections, we are reducing the number of counts available for imaging. This is especially impactful for observations with very low livetime, where pile-up dominates the emission. During TR1, the number of grade 0 counts was very low (2-3 times lower than the number of counts observed in TR3), therefore, to obtain a reliable image of the emission we only performed imaging for TRs 2 and 3. The low energy contours (thermal dominated, yellow) from TR3 were co-aligned with the hot loop, and the resulting shift was applied to contours in TRs 2 and 3.  

We will first focus on the location of the thermal sources. The 131 \AA{} AIA images do not show any significant presence of hot plasma during TR 1, which is consistent with the single 5 MK thermal component in the NuSTAR spectra. The 94 \AA{} channel (bottom panel in Figure \ref{fig:images}) with peak sensitivity to material at about 7 MK \citep[][]{AIA_response}, shows the coronal loop structure that was present before the onset, and is unchanged throughout. Following the first TR, a hot loop structure appears and is present in the 131 \AA{} AIA channel until the start of the flare. The NuSTAR thermal contours in TRs 2 and 3 also remain stationary in the corona, as expected from the additional $>10$ MK thermal component in the spectra.

\begin{figure*}
\centering

\begin{minipage}{0.50\textwidth}

\includegraphics[width=1\textwidth, left]{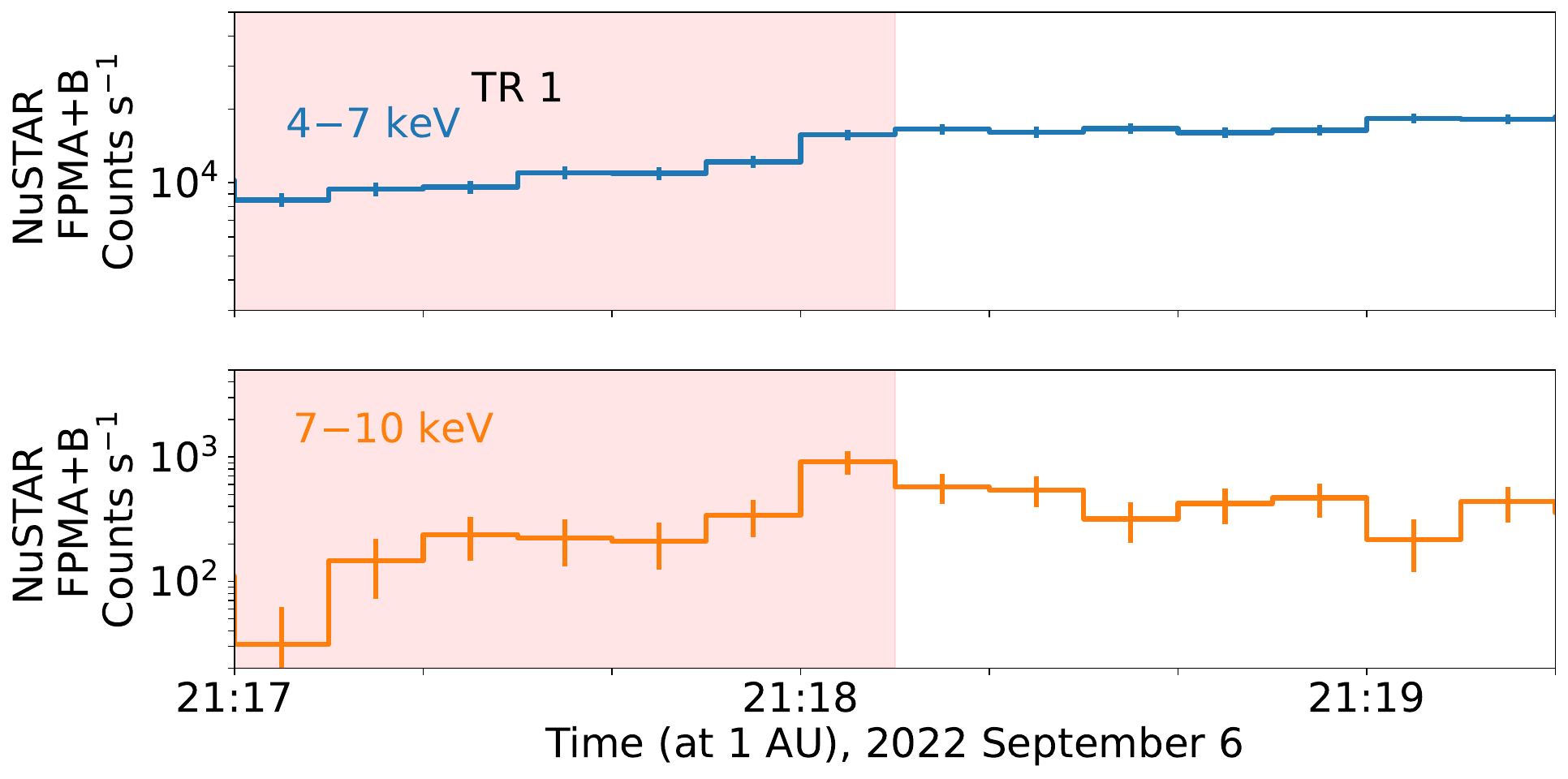}
\end{minipage}

\hspace*{0.06\textwidth}
\begin{minipage}{0.48\textwidth}
    \includegraphics[trim={1.2cm 12.5cm 1.5cm 1.5cm},clip, width=1\textwidth]{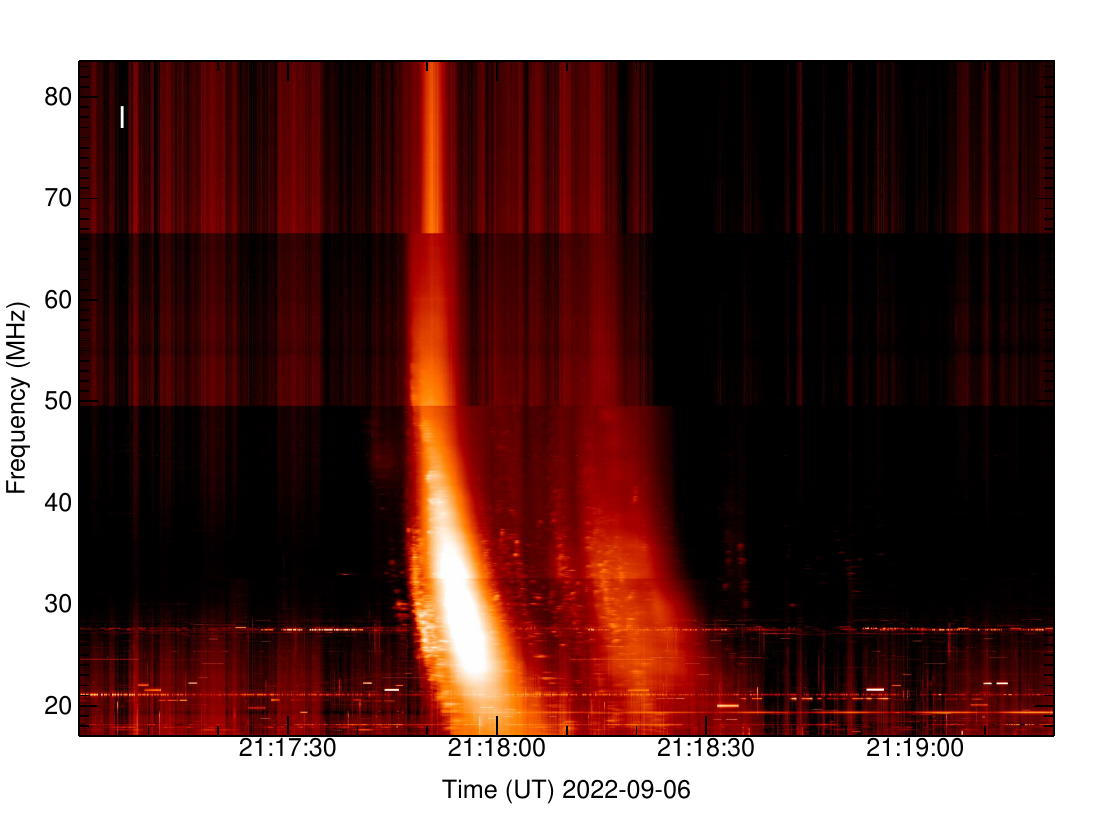}
\end{minipage}

\caption{(Top row) NuSTAR lightcurves from Figure \ref{fig:lightcurves} but limited to time range 1 and part of time range 2. (Bottom row) Series of type III radio bursts observed by the LWA-1 station. These bursts temporally co-align with the end of time range 1. The vertical bands in the upper part of the radio spectrum are interference.}
\label{fig:radio}
\end{figure*}

The nonthermal coronal source was also resolved in the NuSTAR TR3 image (red contours) and appears slightly higher than the brightest part of the thermal source. It appears to be located in the corona above the hot 131 \AA{} AIA loop. Unfortunately, TRs 1 and 2 did not have sufficient number of higher energy counts to form a reliable emission contours. However, from the spectral fits it is clear that there is nonthermal emission present throughout those times. None of the AIA channels show activity during this time that can be linked to the nonthermal source. Furthermore, toward the end of TR1, we also observe a series of type III radio bursts with Long Wavelength Array-1 \citep[LWA-1;][]{LWA_paper} as shown in Figure \ref{fig:radio}. The lack of higher-frequency emission above 250 MHz is consistent with observations of an occulted source. This further confirms that there indeed is electron acceleration during this early phase. Because there is no imaging available for the radio emission, it is hard to constrain its origin; however, it is likely to be connected to the observed NuSTAR event based on timing and occultation. The presence of type III emission also implies that the source has to have access to open or quasi-open field lines. Therefore, we expect that the acceleration occurred behind the limb, and some of the accelerated electrons escaped through any open/quasi-open field lines in the AR. The acceleration mechanism is unknown and hard to constrain due to the lack of available observations from the rear side of the Sun. However, the presence of nonthermal emission both with LWA-1 and NuSTAR could explain the appearance of a hot $\sim$ 11 MK plasma in TR2 and 3, suggesting there might also be downward accelerated electrons that could heat up the chromospheric material that subsequently fills the loop which is seen in AIA. The hot loop from TRs 2 and 3 aligns well with the already present 94 \AA{} AIA loop (as shown by the 131 \AA{} contours overlain on the 94 \AA{} images in Figure \ref{fig:images}), so the observed heating is occurring within the pre-existing coronal loop. Therefore, electron acceleration is likely occurring in a different structure, and the AIA loop is not directly involved in the energy release process - given that the loop remains unchanged and the radio bursts suggest the presence of open field lines. Consequently, NuSTAR is detecting the same electron population as the radio emission which was accelerated either behind the AIA loop or was possibly completely occulted from Earth. Since AIA only sees the top of an occulted loop structure, it is hard to conclude its tilt and height, therefore it is tricky to interpret the location of nonthermal source.

Unfortunately, because of the limiting number of higher energy counts combined with the limited angular resolution of NuSTAR, it is not possible to form any concrete conclusions about the shape of the coronal sources. However, it is clear from the NuSTAR image that there is a clear signature of nonthermal coronal emission near the observed loop.

\section{Analysis of the impulsive flare emission}\label{sec:flare}

\begin{figure*}
\centering
\begin{minipage}{0.27\textwidth}
    \includegraphics[width=1\textwidth]{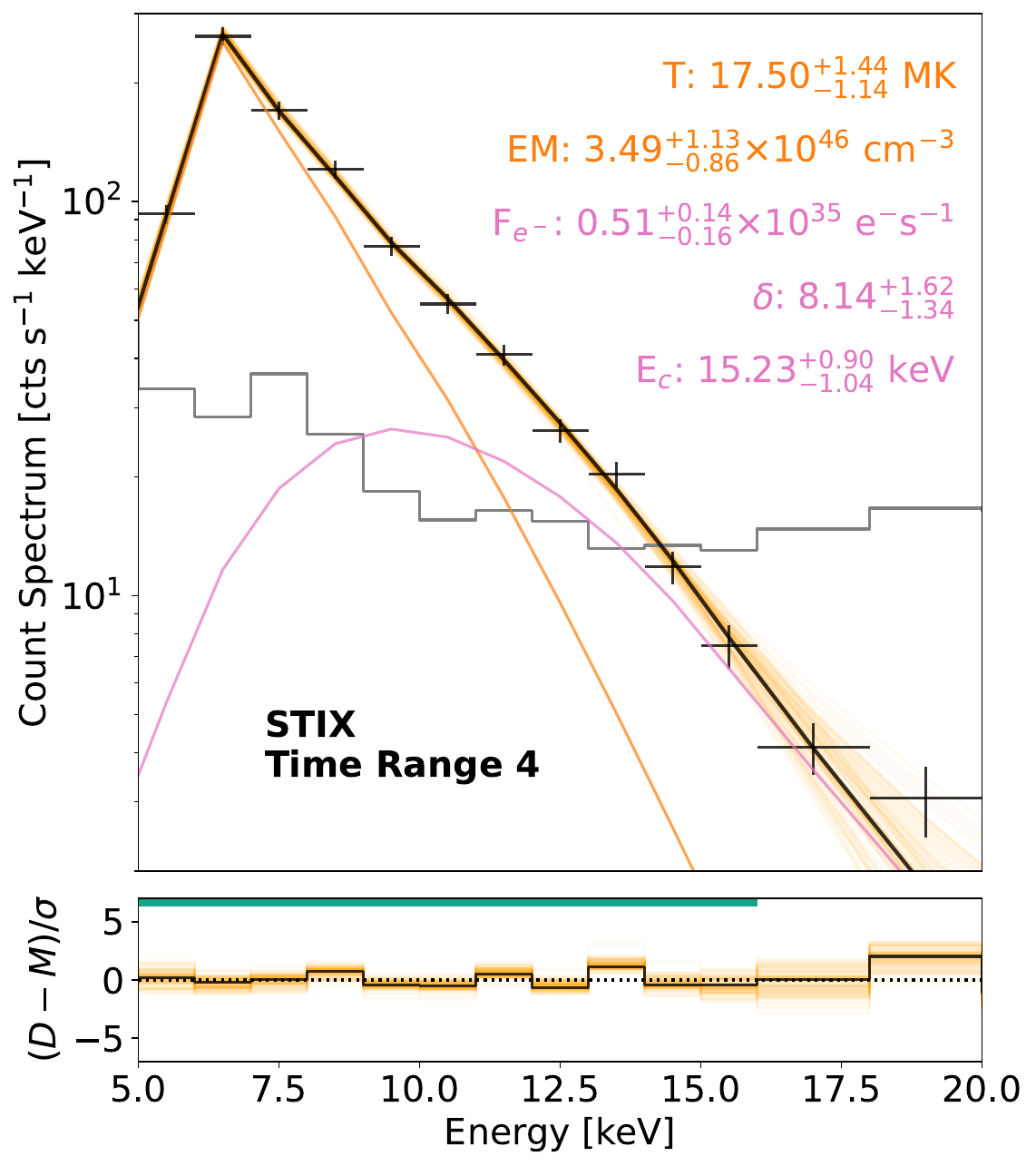}
\end{minipage}
\begin{minipage}{0.36\textwidth}
    \includegraphics[width=1\textwidth]{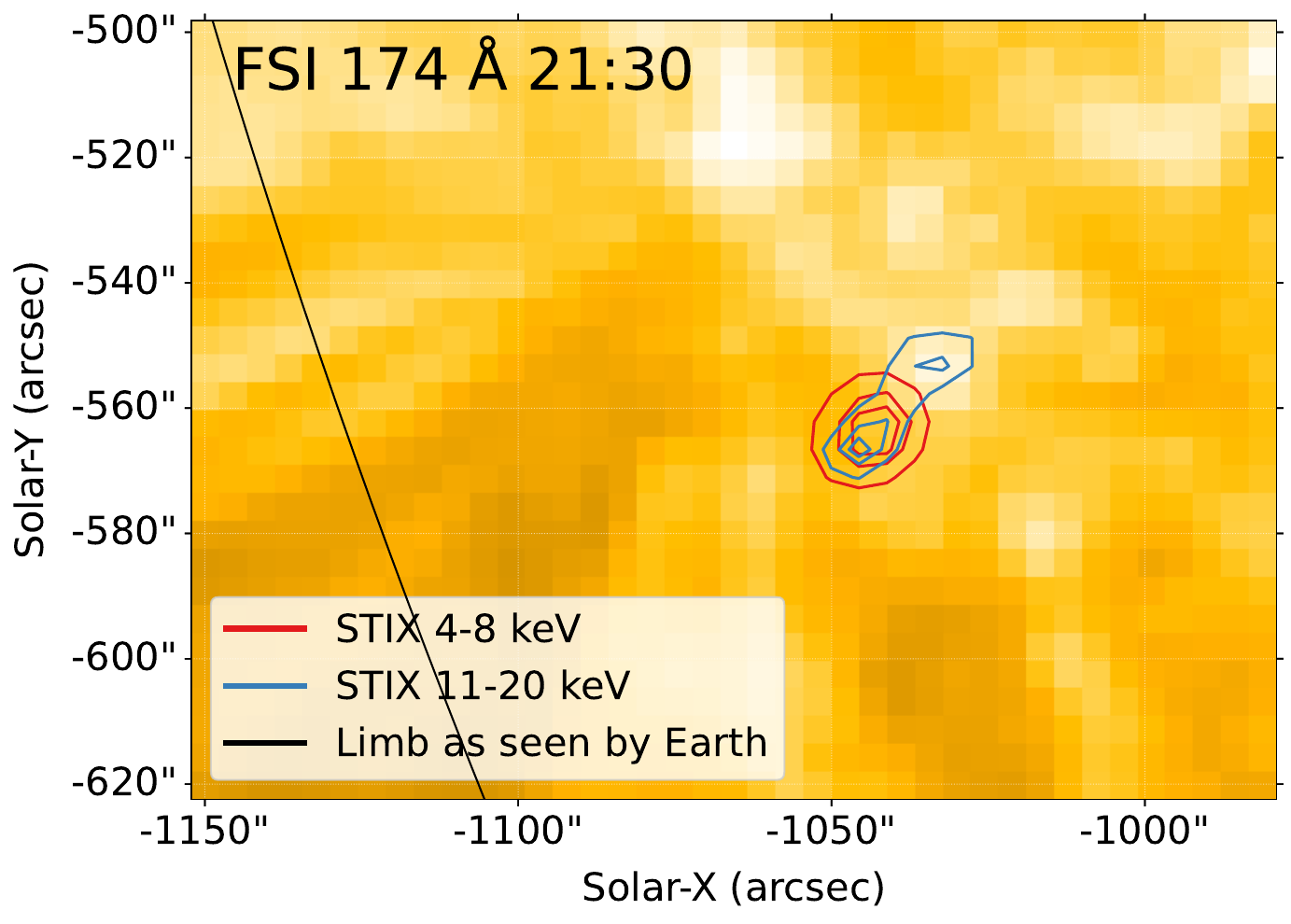}
\end{minipage}
\begin{minipage}{0.30\textwidth}
    \includegraphics[width=1\textwidth]{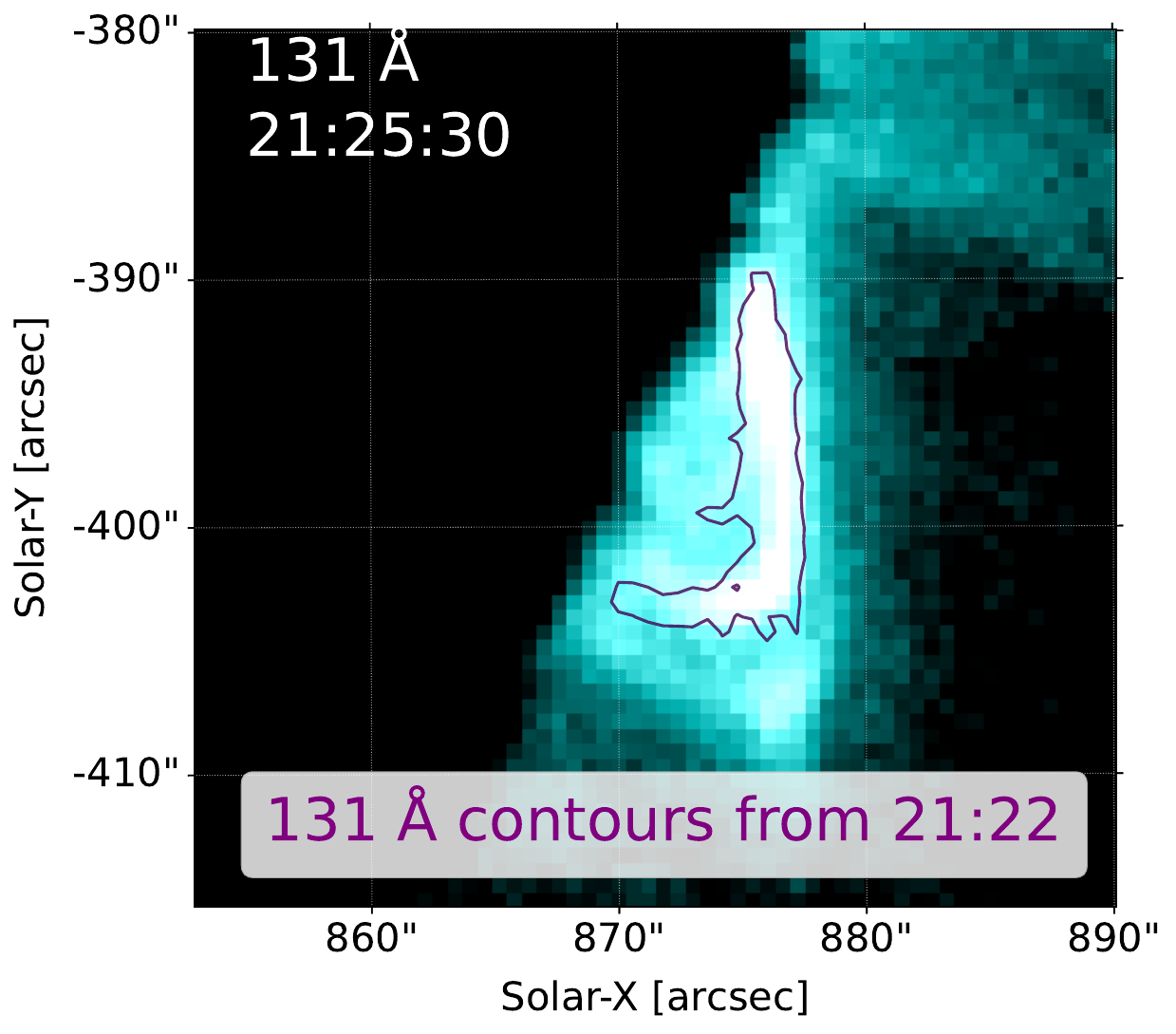}
\end{minipage}

\caption{(Left panel) STIX spectral fit and residuals from time range 4 showing background subtracted spectra, with the backgrounds shown in grey. The fitted model includes thermal (orange lines), thick-target nonthermal (pink) models. The black line indicates the mean MCMC-fitted model parameters with several sample fits shown in yellow. The fitted energy range is shown in the solid green horizontal line. (Middle panel) STIX contours (60, 80 and 90 \%  levels) from TR4 overlain on an FSI image from 21:30. The red contours represent thermally dominated emission and the blue contours are nonthermally dominated emission. (Right panel) 131 \AA{} AIA image from TR4 with 60\% contours from 131 \AA{} image from TR3.}
\label{fig:spectra}
\end{figure*}

The flare was fully observed on-disk by STIX. Thus, we would expect STIX to see the full hot flare loops and footpoints. We fit the integrated STIX spectra from TR 4 highlighted in the lightcurves in Figure \ref{fig:lightcurves}. The spectral fit is shown in the left panel of Figure \ref{fig:spectra} and the best-fit parameters are summarised in Table \ref{tab:spec_params}. The imaged STIX emission is shown in the middle panel in Figure \ref{fig:spectra}. 

The STIX flare spectrum was fitted with thermal and nonthermal thick-target models. We chose to use a thick-target nonthermal model \citep[][]{Brown_1971} as STIX observed the flare to be on-disk and therefore the nonthermal emission should be dominated by the bright footpoints as shown by the higher energy contours in the middle panel in Figure \ref{fig:spectra}.  The $\delta$ from the onset thin-target sources is within the uncertainty range of the $\delta$ from the thick-target model. However, the E$_\textsc{c}$ from the flare footpoints is almost double that of the onset coronal source. This result could suggest that the acceleration mechanism of the onset coronal electron distribution is similar to the one accelerated and stopped in the chromosphere during the flare, with some changes to the acceleration efficiency implied by the E$_\textsc{c}$ increase. This can be due to the acceleration process itself or the change in the plasma conditions. It is also possible that the increase in E$_\textsc{c}$ during the flare is due to a rise in plasma temperature, which would imply that the thermal source is significantly brighter during the flare compared to the onset. Consequently, it becomes more difficult to observe the nonthermal electron distribution at lower energies during the flare. The nonthermal model parameters have higher uncertainties due to the higher energy counts being dominated by background.

STIX imaging from the impulsive phase shows the ``standard" flare configuration with nonthermal footpoints located on either end of a thermal source. Due to the elevated background from SEPs, the higher energy emission has a limited number of counts; therefore, the higher energy contours are marginal compared to the lower energy emission. The image was reconstructed using MEM\_GE \citep[][]{Massa_2020}, with angular resolution of 15 arcsec, and the uncertainty in the location of the reconstructed emission is no better than 10 arcsec. Unfortunately, the absence of EUI images during the impulsive phase of the flare meant that we were unable to get further context from EUV emission on where the STIX contours could originate from, and thus could not improve the precision of their location. 

The lightcurves in Figure \ref{fig:lightcurves} show a similar time profile for both the STIX thermal emission and AIA 131 \AA{}, implying they are both coming from the same structure. The AIA image from this time (TR4) with onset 131 \AA{} loop contours is shown in the right panel in Figure \ref{fig:spectra}. From the overlain onset contours it appears that the same loop structure continues to brighten even during the flare start without a change of its structure, but with some additional heated material. So it appears that AIA might be observing the top of the loop structure, which is fully seen by STIX but with footpoints occulted from the Earth. To examine the location of the STIX emission relative to the AIA flare loops we modeled a semi-circular coronal loop with footpoints anchored at the centers of the nonthermal contours, ensuring that the loop also aligned with the thermal contours. We then re-projected this loop onto the the 131 \AA{} AIA image taken at the time of the flare. A taller loop ($>$25 Mm) or one that is elongated and tilted toward the limb would be required to be anchored at the footpoint location of STIX and with top of the visible above the limb. Due to the lack of high resolution EUV images from the far side of the Sun, it is difficult to determine the configuration of the loop structures. It may be that the scenario is more complicated, with more than one loop structure being heated simultaneously during the flare, so what is seen on AIA and STIX might not be exactly the same structure. However in terms of the energy release site during the onset and flare time, the observations are suggestive of a source somewhere in the corona, possibly (partially) occulted from the Earth. Since the onset NuSTAR electron spectral index $\delta$ matches the one from the STIX footpoints, we speculate that the outwards and downwards accelerated electrons from both the onset and flare times are coming from the same unseen acceleration region. During the onset time NuSTAR and the radio observations revealed nonthermal electrons moving upwards from this location. Then, as the flare progresses, there would be a substantial downwards transport of accelerated electrons, which are seen at the STIX nonthermal footpoints, which would heat the lower atmosphere material and expand upwards, producing the bright sources in AIA and STIX.

\begin{deluxetable*}{lcccccc}
\label{tab:spec_params}
\tabletypesize{\scriptsize}
\tablewidth{0pt} 
\tablecaption{Summary of the best-fit parameters. The first column summarizes the times used for spectrum integration as well as the telescope and time-range number. The remaining columns contain the temperature (T) and emission measure (EM) thermal model parameters and the nonthermal model parameters that include the electron flux F$_{e^{-}}$ (for thick-target), the mean source electron spectrum $<$nVF$>$ (for thin-target), electron spectral index $\delta$ and low-energy cut-off E$_\textsc{c}$.}
\tablehead{
\colhead{Time Range (UT)} & \colhead{\vtop{\hbox{\strut T}\hbox{\strut (MK)}}}& \colhead{\vtop{\hbox{\strut EM }\hbox{\strut ($\times 10^{46} $ cm$^{-3}$)}}} & \colhead{\vtop{\hbox{\strut F$_{e^{-}}$ }\hbox{\strut ($\times 10^{35} $ e$^{-}$s$^{-1}$)}}} & \colhead{\vtop{\hbox{\strut $<$nVF$>$ }\hbox{\strut ($\times 10^{54} $ e$^{-}$cm$^{-2}$s$^{-1}$)}}} & \colhead{$\delta$} & \colhead{\vtop{\hbox{\strut E$_\textsc{c}$ }\hbox{\strut (keV)}}} 
} 
\startdata 
21:15:00 -- 21:17:00 (0, NuSTAR) & $5.64^{+0.07}_{-0.07}$ & $10.10^{+0.80}_{-0.65}$ & --- & --- & --- & ---  \\[0.3cm]
21:17:00 -- 21:18:10 (1, NuSTAR) & $5.65^{+0.10}_{-0.11}$ & $10.75^{+0.91}_{-0.95}$ & --- & $0.38^{+0.10}_{-0.09}$ & $6.23^{+0.98}_{-0.77}$ & $5.47^{+0.53}_{-0.37}$ \\[0.3cm]
21:18:10 -- 21:20:00 (2, NuSTAR) & \vtop{\hbox{\strut $5.95^{+0.20}_{-0.18}$}\hbox{\strut $10.75^{+0.49}_{-0.49}$ }} &
\vtop{\hbox{\strut $8.52^{+1.01}_{-0.90}$}\hbox{\strut $0.20^{+0.08}_{-0.08}$ }} & --- &  $0.37^{+0.09}_{-0.08}$ & $6.92^{+0.94}_{-0.75}$ & $6.25^{+0.23}_{-0.24}$ \\[0.6cm]
21:20:00 -- 21:22:30 (3, NuSTAR) & \vtop{\hbox{\strut $5.99^{+0.14}_{-0.12}$}\hbox{\strut $10.44^{+0.48}_{-0.52}$}} &
\vtop{\hbox{\strut $8.69^{+0.68}_{-0.71}$}\hbox{\strut $0.29^{+0.11}_{-0.09}$ }} & --- & $0.42^{+0.08}_{-0.08}$ & $6.89^{+0.77}_{-0.80}$ & $6.49^{+0.28}_{-0.28}$ \\[0.6cm]
21:25:16 -- 21:25:45 (4, STIX) & $17.50^{+1.44}_{-1.14}$ & $3.49^{+1.13}_{-0.86}$ & $0.51^{+0.14}_{-0.16}$ & --- & $8.14^{+1.62}_{-1.34}$ & $15.23^{+0.90}_{-1.04}$  \\[0.3cm]
\enddata

\end{deluxetable*}

\section{Discussion and Conclusions}\label{sec:discussion}

In this work, we have studied HXR nonthermal emission starting 7 minutes prior to the impulsive energy release of a GOES C-class flare. The onset emission was observed by NuSTAR, while the main flare emission was observed by STIX. The flare was observed to be occulted ($\sim7$\textdegree{} away from the solar limb) for Earth and on-disk for STIX, which allowed us to probe the faint coronal emission with NuSTAR while having the full view of the flare with STIX.

The NuSTAR spectral fitting from the first stage of the onset, lasting $\sim$ 1 minute, shows a sudden appearance of nonthermal thin-target emission but with no detectable increase of the thermal component. In isolation this detection of a weak nonthermal signature before the flare could be seen as marginal given the issues affecting the NuSTAR spectra due to the very low livetime of this observation. However, the simultaneous observations of type III radio emission with LWA-1 that come from an occulted source (due to the lack of higher frequency emission) strengthen the case for the NuSTAR nonthermal detection.

After the start of the onset of electron acceleration, both NuSTAR and the 131 \AA{} AIA channel observed a hot plasma at 10-11 MK temperatures in a coronal loop that was observed during the preflare time in the 94 \AA{} AIA channel. This hot plasma was observed at approximately constant temperatures for almost 5 minutes, which could point toward the hot-onset scenario. The NuSTAR observations during TR1-3 match the expectation from the GOES observations of hot onsets, except that the emission measure does not show an increase between TR2 and TR3, the ``horizontal branch" noted by \citet{Hugh_onset}. GOES did not detect the hot onset in this event due to the emission measure (found with NuSTAR) being an order of magnitude below its sensitivity threshold. In addition to the hot thermal plasma, during TRs 2-3 NuSTAR also observed a thin-target coronal source near the loop. Throughout the whole onset-stage, the nonthermal sources have the same $\delta$ and show a small increase in $E_{\text{c}}$. Unlike the NuSTAR thermal source, we could not find a detectable change in the AIA observations that coincides with the NuSTAR nonthermal source. The fact that the loop structure remained unchanged throughout the onset period, combined with temporally co-aligned type III radio bursts from an occulted source, indicates that the acceleration occurred behind the AIA loops, and likely entirely behind the limb. Therefore, both NuSTAR and LWA-1 are observing the same upward-accelerated electron population, while the downward-accelerated electrons result in chromospheric heating that subsequently fills the AIA loop in TRs 2 and 3. A similar scenario where preflare plasma heating is caused by electron acceleration was previously reported by \citet{Siarkowski_2009}. However, given the ambiguity in the physical extent of the sources, we cannot determine their full energy content and conclusively confirm this scenario.

This 7 minute long onset stage observed with NuSTAR was followed by the flare start. The spectral fitting and imaging with STIX from the impulsive phase show the presence of thermal emission from hot flare loops and nonthermal thick-target footpoint emission. The $\delta$ of the thick-target emission is within the uncertainties of the NuSTAR onset coronal thin-target source, while the thick-target $E_{\text{c}}$ is more than double the coronal onset $E_{\text{c}}$. The consistent $\delta$ could be a signature of a similar acceleration mechanism throughout the onset and flare times, originating from the same (possibly occulted) acceleration region. The STIX nonthermal footpoints are fully occulted from Earth, but the observed AIA limb structure may be the top of the thermal loops seen by STIX due to their similar time profiles. This would require either a very tall, or elongated and tilt loop structure. However, given the lack of high resolution EUV images from behind the limb, it is difficult to conclude the specific configuration of these loops. An alternative scenario could be the AIA and STIX thermal sources being different loop structures that are simultaneously heated by the flare accelerated electrons.

Given the uncertainty in the exact configuration of this onset and flare energy release, modelling work is required in the future to determine whether the observed heating signatures are consistent with the onset nonthermal source. This might be aided by additional observations available from the Gamma-ray Burst Monitor (GBM) on board the Fermi Gamma-ray Space Telescope \citep[Fermi;][]{Meegan_2009}, which gives us the opportunity to study coronal emission from the flare time and provide a further link between flare coronal emission and the onset coronal emission from NuSTAR. However, the GBM emission during TR4 is very faint and mostly dominated by background; therefore, the spectrum is limited to energies up to $\sim$ 25 keV. Due to the uncertainty in the GBM response at low energies, combined with the low count rate, we would require a more systematic cross-calibration study of such faint events with other X-ray instruments prior to performing the spectral analysis of this flare.

Although nonthermal preflare coronal emissions have been previously reported in \citet{Asai_2006}, this analysis shows the first detailed evidence of the spatial, temporal, and spectral evolution of such onset nonthermal coronal sources. This work indicates the opportunities that come from solar flare studies using multiple viewpoints, one catching an occulted flare and the purely coronal emission, the other the on-disk view of the lower atmosphere footpoints and coronal emission \citep[e.g.][]{Ryan2024, Krucker_HXI}.

\begin{acknowledgments}
    NB acknowledges support from the UK’s Science and Technology Facilities Council (STFC) doctoral training grant (SST/X508391/1). IGH acknowledges support from STFC grant ST/X000990/1. SW acknowledges support for basic research from AFOSR grant LRIR 23RVCOR003. SK is supported by Swiss PRODEX grant for STIX.

    This paper made use of data from the NuSTAR mission, a project led by the California Institute of Technology, managed by the Jet Propulsion Laboratory, funded by the National Aeronautics and Space Administration. Throughout the NuSTAR analysis, we made use of the NuSTAR Data Analysis Software (NUSTARDAS) jointly developed by the ASI Science Data Center (ASDC, Italy), and the California Institute of Technology (USA). Solar Orbiter is a space mission of international collaboration between ESA and NASA, operated by ESA. The STIX instrument is an international collaboration between Switzerland, Poland, France, Czech Republic, Germany, Austria, Ireland, and Italy. The STIX analysis was done using the SolarSoft IDL distribution (SSW) version 0.5.1 from the IDL Astronomy Library. Construction of the LWA has been supported by the Office of Naval Research and by the AFOSR. Support for operations and continuing development of the LWA1 is provided by the Air Force Research Laboratory and the National Science Foundation. This research also made use of the SunPy open source software package \citep{sunpy_2020} and sunkit-spex, a Python solar X-ray spectral fitting package under the SunPy development. We would like to thank the anonymous referee for their helpful feedback.
\end{acknowledgments}

\section*{Data Availability}
NuSTAR data can be obtained from the NuSTAR Master Catalogue\footnote{\href{https://heasarc.gsfc.nasa.gov/docs/nustar/nustar_archive.html}{\nolinkurl{https://heasarc.gsfc.nasa.gov/docs/nustar/nustar_archive.html}}} with OBSID 90810204001. STIX data can be accessed through Solar Orbiter Archive\footnote{\href{https://soar.esac.esa.int/soar/}{\nolinkurl{https://soar.esac.esa.int/soar/}}} with UID 2209064468. LWA1 data can be access through the LWA data archive\footnote{\href{https://lda10g.alliance.unm.edu/ldadb/observations/search/ }{\nolinkurl{https://lda10g.alliance.unm.edu/ldadb/observations/search/ }}}.


\bibliography{references}{}
\bibliographystyle{aasjournalv7}



\end{document}